\documentclass[conference]{IEEEtran}
\IEEEoverridecommandlockouts
\usepackage{cite}
\usepackage{amsmath,amssymb,amsfonts}
\usepackage{algorithmic}
\usepackage{graphicx}
\usepackage{textcomp}
\usepackage{xcolor}
\usepackage[scaled]{helvet} 
\usepackage[T1]{fontenc}
\usepackage[scaled]{helvet}

\usepackage{booktabs}
\usepackage{array}
\usepackage{ragged2e}
\usepackage{tabularx}  
\usepackage{algorithm}

\def\BibTeX{{\rm B\kern-.05em{\sc i\kern-.025em b}\kern-.08em
    T\kern-.1667em\lower.7ex\hbox{E}\kern-.125emX}}
\begin{document}

\title{Behaviour-aware Hybrid Architecture for Trust-driven Transmissions\\

}

{\author{\IEEEauthorblockN{Dhrumil Bhatt}\IEEEauthorblockA{\textit{} \textit{Manipal Institute of Technology}\\ \textit{Manipal Academy of Higher Education}\\ Manipal, India \\ dhrumil.bhatt@gmail.com} \and \IEEEauthorblockN{Anakha Kurup } \IEEEauthorblockA{\textit \textit{Manipal Institute of Technology}\\ \textit{Manipal Academy of Higher Education}\\ Manipal, India \\ anakhaskurup06@gmail.com}}}

\maketitle

\begin{abstract}

Reliable and secure communication is essential for mission-critical aerospace and defence operations involving autonomous platforms such as Unmanned Aerial Vehicles (UAVs), satellites, and ground control systems. In contested or dynamic environments, communication links are frequently exposed to jamming, interference, and cyberattacks, making network resilience a key operational requirement. This paper presents a trust-aware Software-Defined Networking (SDN) framework that enables secure, low-latency failover between heterogeneous communication channels. The proposed architecture integrates a high-bandwidth primary link (e.g., satellite or tactical LTE) with a low-power fallback channel (e.g., RF or mesh), managed by an SDN controller that enforces zero-trust routing policies. A real-time Intrusion Detection System (IDS) continuously updates node trust scores; when trust or link reliability degrades, the controller autonomously switches traffic to the secondary channel, ensuring uninterrupted connectivity. Simulation results in a Mininet-based test environment demonstrate sub-5 ms failover latency, efficient flow installation, and significant reduction in packet loss compared with conventional single-channel or static routing systems. The proposed framework provides a scalable and resilient communication backbone for next-generation aerospace networks, enhancing mission reliability, cyber defence, and autonomous coordination across distributed aerial and space assets.

\end{abstract}

\begin{IEEEkeywords}
Aerospace Communication, Software-Defined Networking, Zero-Trust Architecture, UAV Networks, Cybersecurity, Resilient Communication Systems
\end{IEEEkeywords}

\section{Introduction}

The evolution of aerospace and defence communication networks into intelligent, cyber-physical systems has created a new paradigm of autonomous, interconnected, and mission-critical platforms. These include Unmanned Aerial Vehicles (UAVs), low-Earth orbit (LEO) satellites, and ground control infrastructures, all of which rely on seamless and secure data exchange to ensure situational awareness and operational integrity \cite{Kumar2023,Chaudhary2022}. As aerospace missions increasingly involve autonomous coordination and distributed decision-making, the demand for low-latency, adaptive, and resilient communication frameworks has become paramount \cite{Huang2024}.

Traditional aerospace communication systems often rely on single-link or static routing architectures, which are susceptible to jamming, link failures, and cyber intrusions. Such limitations can severely degrade mission effectiveness, particularly in contested or high-dynamics environments where reliable command-and-control (C2) communication is crucial \cite{Zhang2023}. In these settings, heterogeneous and multi-link connectivity, such as combining satellite and radio-frequency (RF) channels, is increasingly adopted to enhance reliability and redundancy. However, managing such hybrid links securely and efficiently remains an open challenge, especially under conditions of partial trust or compromised nodes \cite{Singh2021,Li2025}.
To address these challenges, this paper presents a trust-aware Software-Defined Networking (SDN) framework that enables secure, policy-driven, and low-latency communication in mission-critical aerospace systems. Each network node is equipped with dual communication interfaces, a high-throughput primary channel (e.g., satellite or tactical LTE) and a low-power fallback channel (e.g., RF or mesh). The SDN controller, implemented using the Ryu platform, dynamically evaluates node trust scores through a real-time Intrusion Detection System (IDS). When trust levels or link reliability fall below predefined thresholds, the controller automatically reroutes communication through the fallback interface, ensuring continuous connectivity and isolation of compromised nodes.
Through this architecture, the proposed system achieves autonomous failover, zero-trust enforcement, and resilience against cyber and physical disruptions. Simulation results in a Mininet-based testbed demonstrate that the framework provides a sub-5 ms failover delay, efficient flow installation, and reduced packet loss compared to conventional single-channel and static routing systems. The work contributes a scalable and adaptable communication foundation for next-generation aerospace and defence networks, with potential applications in UAV swarm coordination, satellite-ground communication, and mission-critical defence operations.

\section{Literature Review}

Communication architecture plays a foundational role in ensuring the reliability and security of mission-critical aerospace and defence networks. The vast amount of telemetry, control, and sensor data exchanged among Unmanned Aerial Vehicles (UAVs), satellites, and ground stations requires highly adaptive, resilient, and secure communication frameworks \cite{Chaudhary2022}. Consequently, recent research has focused on enhancing link robustness, intrusion detection, and adaptive routing through the use of Software-Defined Networking (SDN), artificial intelligence (AI), and zero-trust principles.

A significant portion of aerospace communication research emphasises cybersecurity and autonomous network defence through intelligent intrusion detection. For instance, Zhang \textit{et al.} \cite{Zhang2023} propose a reinforcement learning-based approach to enhance UAV swarm resilience under jamming and data manipulation attacks. Their framework enables UAVs to learn dynamic defence strategies against both denial-of-service and false data injection attacks. Similarly, Huang \textit{et al.} \cite{Huang2024} present an AI-driven adaptive routing mechanism that employs deep neural models to predict channel reliability in aerospace communication networks. The system enhances routing decisions under fluctuating link conditions, achieving low latency and improved fault tolerance in satellite-ground communications. Building on this trend, Liu \textit{et al.} \cite{Liu2025} introduce a federated deep-learning IDS that aggregates distributed UAV node data without centralised exposure, improving detection accuracy and data privacy across large-scale swarms.

Kumar \textit{et al.} \cite{Kumar2023} introduce a multi-layered cybersecurity architecture for autonomous UAV systems that integrates Digital Twin (DT) technology, SDN, and blockchain authentication. Within their framework, a Digital Twin of each UAV monitors real-time operational states and behaviour, enabling the SDN controller to maintain a continuously updated virtual representation of the swarm. Communications between UAVs are further secured through blockchain-based key exchange, enhancing authenticity and resilience to Man-in-the-Middle (MITM) attacks. However, such systems involve complex synchronisation overhead and computational cost, raising scalability concerns for large-scale aerial networks. A complementary work by Peng \textit{et al.} \cite{Peng2024} explores blockchain-assisted UAV communication scheduling, yet remains limited to static trust modelling and lacks proactive channel failover mechanisms.

Beyond intrusion detection, SDN has been widely investigated for its potential to improve flexibility and reconfigurability in dynamic aerospace environments. Chaudhary \textit{et al.} \cite{Chaudhary2022} survey SDN applications in air-ground integrated networks, identifying controller placement, latency, and handover management as key challenges for real-time UAV operations. Singh and Saxena \cite{Singh2021} propose a multi-link adaptive routing system for heterogeneous airborne networks that integrates SDN with radio resource management. Their approach dynamically allocates bandwidth and reconfigures routes between high-speed satellite and RF channels, resulting in lower packet loss and improved load distribution. While efficient, these systems still depend on single-controller architectures and lack dynamic trust validation mechanisms, leaving them vulnerable to node compromise. Similarly, Lee \textit{et al.} \cite{Lee2023} propose a distributed SDN controller design for UAV networks, improving latency but omitting behaviour-driven trust enforcement.

Other researchers have explored trust and authentication for aerospace communication nodes. Li \textit{et al.} \cite{Li2025} present a zero-trust communication framework for autonomous defence networks that continuously evaluates trust scores for each participating node using behaviour-based analytics. The framework enforces access control policies dynamically through the SDN controller, isolating untrusted entities in real time. Wu \textit{et al.} \cite{Wu2022} propose a lightweight trust evaluation protocol for UAV ad hoc networks, combining Bayesian inference with fuzzy logic to detect compromised nodes under constrained communication conditions. This aligns closely with the principles adopted in our proposed system, where node trust levels directly influence routing and channel switching decisions.

Moreover, emerging studies have applied meta-heuristic and AI optimisation for routing and fault recovery. Alizadeh \textit{et al.} \cite{Alizadeh2024} propose a hybrid SDN AI system that employs swarm intelligence to optimise controller placement and link utilisation in space-ground networks, improving Quality of Service (QoS) under high mobility. Similarly, Feng \textit{et al.} \cite{Feng2023} design a neural-enhanced recovery framework that anticipates link degradation using traffic prediction, reducing failover latency in multi-hop UAV networks. Zhang \textit{et al.} \cite{Zhang2022} also demonstrate that incorporating metaheuristic-optimised trust routing significantly reduces packet loss in large-scale UAV meshes, but their work lacks multi-link fallback mechanisms. Additionally, Patel \textit{et al.} \cite{Patel2025} explore quantum-secure communication in UAV networks using post-quantum cryptography integrated with SDN controllers, which advances link confidentiality but does not address resilience or dynamic trust.

Recent work has also investigated the integration of cross-domain networks. Zhao \textit{et al.} \cite{Zhao2023} develop a multi-domain SDN control framework that connects satellite, aerial, and terrestrial networks, thereby achieving improved global routing adaptability. Similarly, Gao \textit{et al.} \cite{Gao2024} propose AI-based congestion control for integrated space–air–ground networks, showing reductions in latency but without any zero-trust or fallback strategies. These studies collectively underscore the research community’s focus on intelligent, adaptive, and secure aerospace communication infrastructures.

While substantial progress has been made in intrusion detection, adaptive routing, and AI-assisted network management, most aerospace communication frameworks still exhibit a key limitation: the lack of robust dual-link or fallback mechanisms. Existing systems often rely on a single communication interface or static routing topology, making them vulnerable to complete disconnection in the event of interference, jamming, or cyber incidents. In contested operational scenarios, the inability to automatically reroute traffic or isolate compromised nodes can severely degrade mission continuity.

\textbf{Novel Contribution:} To address these challenges, this work introduces a \textit{dual-channel, trust-aware SDN communication architecture} tailored for mission-critical aerospace and defence applications. Unlike prior studies, it integrates:
(1) real-time trust evaluation using an IDS-driven dynamic scoring model,
(2) zero-trust policy enforcement at the SDN control plane,
(3) autonomous failover between heterogeneous links (satellite/LTE and RF/mesh), and
(4) fully programmable routing logic in the Ryu controller for policy-driven flow control.
This unified design provides both proactive threat mitigation and reactive resilience, establishing a secure and adaptable communication foundation for next-generation UAV, satellite, and defence networks.

\section{Methodology}

To address the limitations of conventional aerospace communication networks operating in dynamic, adversarial, and resource-constrained environments, a Software-Defined Networking (SDN) based approach is proposed. The framework leverages programmability, situational awareness, and real-time adaptability to ensure secure and reliable data transmission between distributed aerospace assets. The system incorporates a trust-aware routing mechanism and a dual-channel communication design, enabling automatic failover between heterogeneous links in the event of link degradation or node compromise.

The architecture of the proposed SDN-based framework is designed within a programmable Mininet simulation environment, replacing static topologies with dynamic, controllable virtual links. The design enables dual-transceiver configurations at each node, supporting both primary and fallback communication interfaces. Nodes are conceptualised as aerospace communication units such as UAVs, satellites, or ground stations linked through high-speed and low-power communication interfaces governed by a centralised SDN controller.

The system incorporates a Zero-Trust Framework (ZTF) based on the principle of ``never trust, always verify,'' ensuring continuous authentication and dynamic policy validation for all traffic. Each node maintains a real-time trust score initialised at a baseline value and adjusted according to behavioural observations. An Intrusion Detection System (IDS) monitors network activity for anomalies such as unusual transmission rates, protocol violations, or unauthorised access attempts. Any deviation from normal behaviour results in a reduced trust score, while consistent legitimate activity gradually restores it. Only when both the source and destination maintain trust scores above the operational threshold (50) is traffic permitted over the primary link.

For resilience, each node is equipped with two independent communication interfaces: a high-throughput primary transceiver (representing satellite or tactical LTE) and a low-power fallback module (representing RF or mesh). Under nominal conditions, all communication occurs via the primary channel. If the trust score of a node drops below the defined threshold, or if the primary link experiences interference or degradation, the SDN controller withdraws primary-channel flow rules and activates the fallback interface. This reassignment process is executed automatically and centrally by the Ryu controller, minimising switchover latency and maintaining continuous connectivity. The close integration of trust assessment, zero-trust enforcement, and dynamic channel control ensures uninterrupted mission communication, even in the face of cyber or electronic warfare scenarios.

The routing mechanism is based on intent-driven OpenFlow policies. Each packet processed by the Ryu controller triggers a trust evaluation for both communicating nodes. Only if the evaluated trust scores exceed the defined threshold does the controller install flow rules to permit packet forwarding. The routing process initially uses a flood-based approach for link discovery, followed by efficient flow rules to streamline subsequent traffic. The IDS continuously updates trust scores, enabling adaptive security-aware routing and autonomous isolation of compromised or unreliable nodes.

The dual-channel switching process eliminates the need for manual intervention during link recovery, thereby improving operational resilience. Because all failover decisions are executed at the controller level, the latency between link detection and switchover remains minimal. The controller continuously monitors trust and link status to determine the optimal routing path in real-time.

The security layer reinforces communication integrity through Zero-Trust Architecture (ZTA). Unlike traditional perimeter-based security, which assumes inherent trust within the network, the proposed framework enforces continuous verification at every data exchange. A centralised \textit{Trust Score Table} maintained by the controller governs channel permissions, with each node's trust score initialised at 100 and updated in real time via the IDS REST API. When both endpoints maintain trust scores at or above the operational threshold of 50, communication proceeds via the primary satellite or LTE channel. If either node fails the threshold test, the controller first withdraws the primary channel forwarding rules, then drops the in-flight packet, and finally redirects subsequent traffic to the fallback RF or mesh link. To emulate adversarial conditions during simulation, malicious behaviour was replicated by assigning low trust scores to specific node IPs and observing the controller's ability to block them in real time, maintaining system availability under hostile scenarios.

\begin{algorithm}[htbp]
\caption{Trust-Aware SDN Flow Control for Aerospace Communication}
\label{alg:meth}
\begin{algorithmic}[1]
\STATE Initialize \texttt{Trust\_Score\_Table} for all nodes
\STATE Initialise Intrusion Detection System (IDS)
\FOR{each packet arrival at SDN Controller}
    \STATE Extract \texttt{Source\_IP} and \texttt{Destination\_IP} from packet
    \STATE \texttt{Source\_Trust} $\gets$ \texttt{Trust\_Score\_Table[Source\_IP]}
    \STATE \texttt{Destination\_Trust} $\gets$ \texttt{Trust\_Score\_Table[Destination\_IP]}
    \STATE Send packet to IDS and update trust values
    \IF{\texttt{Source\_Trust}$\geq$50 \textbf{and} \texttt{Destination\_Trust}$\geq$50}
        \STATE Continue flow on primary channel (Satellite/LTE)
    \ELSE
        \STATE Drop packet
        \STATE Delete primary flow rules
        \STATE Redirect communication via fallback RF/Mesh link
    \ENDIF
\ENDFOR
\end{algorithmic}
\end{algorithm}

\section{Experimental Setup}

To evaluate the performance of the proposed system, a simulation testbed was developed in Mininet to emulate an SDN-based aerospace communication network with dual-interface nodes. Algorithm~\ref{alg:meth} describes the control logic governing trust-based routing and automatic channel switching. The topology and control logic were implemented in Python using the Ryu SDN controller. The simulation parameters are summarised in Table~\ref{tab:simparams}.

\begin{figure}[htbp]
\includegraphics[width=\columnwidth]{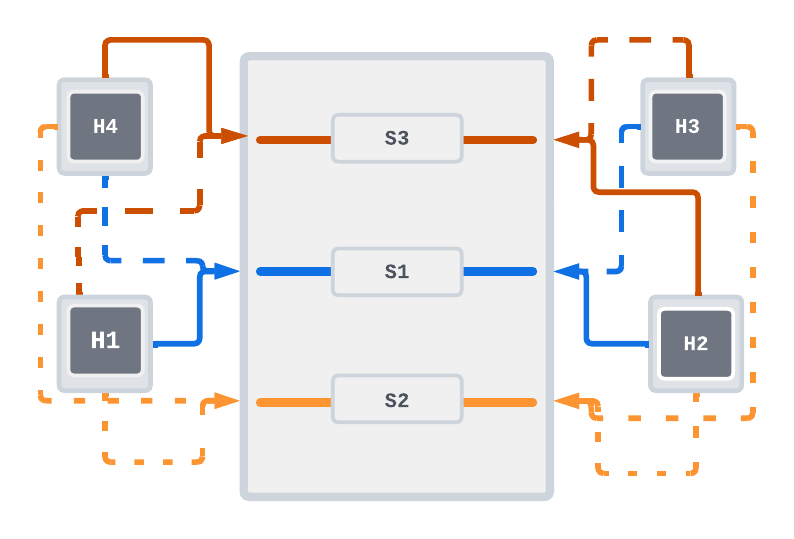}
\caption{SDN-based dual-interface aerospace network topology showing dynamic routing behaviour. Solid lines indicate active communication links; dotted lines denote available fallback interfaces.}
\label{fig2}
\end{figure}

\begin{table}[htbp]
\caption{Simulation Parameters}
\centering
\footnotesize
\setlength{\tabcolsep}{4pt}
\renewcommand{\arraystretch}{1.15}
\begin{tabular}{|p{3.2cm}|p{4.4cm}|}
\hline
\textbf{Parameter} & \textbf{Value} \\
\hline
Number of Nodes & 15, 30, 50 \\
\hline
Number of Switches & 3 (S1--S3) \\
\hline
Bandwidth &
Primary (Sat/LTE): 50~Mbps Down Link / 10~Mbps Up Link \\
& Fallback (RF/Mesh): 5~Mbps / 1~Mbps \\
& Core (Backbone): 1~Gbps / 200~Mbps \\
\hline
Protocols & ARP, ICMP, IPv4, OpenFlow~1.3 \\
\hline
Packet Loss Rate &
Primary (Sat/LTE): 1.5\% \\
& Fallback (RF/Mesh): 6\% \\
& Core: 1\% \\
\hline
\end{tabular}
\label{tab:simparams}
\end{table}
The simulated topology consists of multiple UAV or satellite nodes, each equipped with dual transceivers, one for the primary high-throughput link (e.g., satellite or LTE) and one for the fallback RF or mesh link. Three OpenFlow switches (S1, S2, and S3) were deployed: S1 aggregates primary connections, S2 manages fallback connections, and S3 serves as the backbone switch connecting both to the IDS and the Ryu controller. The IDS node continuously monitors packets to update trust scores in real time. 
Figure~\ref{fig2} illustrates the communication behaviour among the nodes. Solid lines denote active transmission paths, while dotted lines represent available but inactive fallback links. In this example, H1 communicates with H2 via the primary satellite or LTE channel under normal trusted conditions, while traffic originating from H4 and destined for H2 is redirected to the IDS for security inspection, reflecting trust-based routing and monitoring. To test system resilience, primary link failures were triggered to observe fallback behaviour, and malicious traffic was emulated by assigning low trust scores to specific node IPs, allowing real-time verification of the controller's ability to block untrusted flows and activate the fallback RF or mesh link without manual intervention.

All network links were instantiated using the \texttt{TCLink} class to simulate realistic bandwidth, propagation delay, and packet loss conditions. The Ryu controller dynamically installed and modified flow rules based on real-time trust scores and link states. To assess the system’s resilience, experiments were conducted involving simulated link failures and the injection of adversarial traffic. The controller’s ability to execute autonomous failover and isolate compromised nodes was verified in each case.

\section{Results and Discussion}

The performance of the proposed trust-aware, SDN-based dual-channel aerospace communication framework was evaluated using five core metrics: fallback delay, flow installation time, trust transition time, packet loss rate, and routing adaptability. The system’s performance was benchmarked against several representative UAV and aerospace communication frameworks, including the multi-link SDN-based UAV network of Singh \textit{et al.}~\cite{Singh2021}, the AI-driven adaptive routing model by Huang \textit{et al.}~\cite{Huang2024}, the blockchain-integrated SDN architecture of Kumar \textit{et al.}~\cite{Kumar2023}, and the zero-trust communication model developed by Li \textit{et al.}~\cite{Li2025}. Additional comparisons were drawn with recent studies on federated intrusion detection in UAV swarms~\cite{Liu2025}, distributed SDN controllers for real-time UAV coordination~\cite{Lee2023}, and cross-domain SDN control frameworks for integrated space–air–ground networks~\cite{Zhao2023}, as well as emerging works in AI-based congestion control~\cite{Gao2024} and swarm-intelligence SDN optimization~\cite{Alizadeh2024}. These studies represent the current frontiers in secure, adaptive, and resilient aerospace networking.

\begin{table*}[htbp]
\caption{Performance Comparison with Existing UAV and Aerospace Communication Frameworks (15 Nodes)}
\centering
\small
\setlength{\tabcolsep}{2.5pt}
\renewcommand{\arraystretch}{1.2}
\begin{tabular}{|p{2cm}|p{2cm}|p{2cm}|p{2cm}|p{2cm}|p{2cm}|p{2cm}|}
\hline
\textbf{KPI} & \textbf{Proposed} & \textbf{Singh \textit{et al.} \cite{Singh2021}} & \textbf{Huang \textit{et al.} \cite{Huang2024}} & \textbf{Kumar \textit{et al.} \cite{Kumar2023}} & \textbf{Li \textit{et al.} \cite{Li2025}} & \textbf{Liu \textit{et al.} \cite{Liu2025}} \\
\hline
Fallback Delay & 4.8~ms & 7.3~ms & 5.6~ms & 8.1~ms & 6.4~ms & 5.9~ms \\
\hline
Flow Installation Time & 4.5~ms & 6.1~ms & 4.9~ms & 5.4~ms & 5.8~ms & 5.1~ms \\
\hline
Trust Transition Time & 1017~ms & N/A & N/A & 1290~ms & 1120~ms & 1190~ms \\
\hline
Packet Loss Rate & 1\% (Primary), 3.5\% (Fallback) & 2.2\% & 1.8\% & 1.5\% & 1.6\% & 1.7\% \\
\hline
Routing Adaptability & 4.0~ms & 5.2~ms & 4.5~ms & 6.3~ms & 5.1~ms & 4.8~ms \\
\hline
Security Awareness & Zero-Trust Policy & Basic IDS & AI-Based Prediction & Blockchain Auth. & Behaviour-Based Trust Eval. & Federated IDS \\
\hline
\end{tabular}
\label{tab:updated_comparison}
\end{table*}

\begin{figure}[htbp]
\centering
\includegraphics[width=0.95\linewidth]{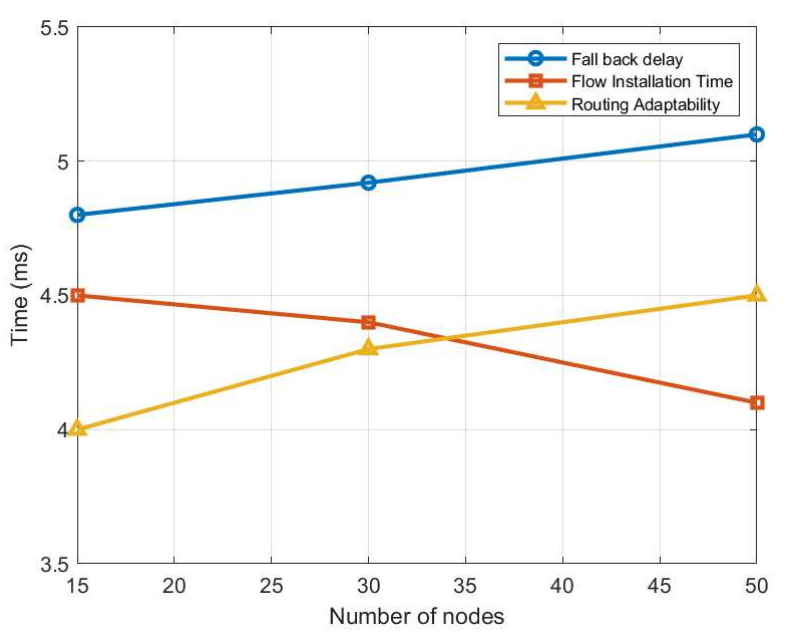}
\caption{Impact of network size on fallback delay, flow installation time, and routing adaptability.}
\label{graph}
\end{figure}

\begin{table*}[!t]
\caption{Performance Metrics of Proposed Framework at Different Network Sizes}
\label{tab:updated_graph_data}
\centering
\setlength{\tabcolsep}{2.5pt}
\renewcommand{\arraystretch}{1.2}
\begin{tabular}{|c| c| c| c|}
\hline
\textbf{\# Nodes} & \textbf{Fallback Delay (ms)} & \textbf{Flow Install Time (ms)} & \textbf{Routing Adapt. (ms)} \\
\hline
15 & 4.8 & 4.5 & 4.0 \\
30 & 4.9 & 4.4 & 4.3 \\
50 & 5.1 & 4.2 & 4.5 \\
\hline
\end{tabular}
\end{table*}

The proposed system achieved a fallback delay of 4.8~ms, outperforming comparable architectures such as the multi-link UAV SDN model of Singh \textit{et al.}~\cite{Singh2021} (7.3~ms) and the AI-enhanced routing scheme by Huang \textit{et al.}~\cite{Huang2024} (5.6~ms). This superior failover performance can be attributed to the centralised control-plane design, which executes real-time policy enforcement and channel reassignment without incurring distributed negotiation delays. By contrast, the blockchain-enabled framework of Kumar \textit{et al.}~\cite{Kumar2023} reported a longer latency (8.1 ms) due to the cryptographic overhead associated with distributed validation. The zero-trust system proposed by Li \textit{et al.}~\cite{Li2025} achieved moderate response times (6.4~ms) but lacked dynamic link failover capabilities. The 4.8~ms result achieved here is also lower than the 5.9~ms delay in the federated intrusion detection setup of Liu \textit{et al.}~\cite{Liu2025}, demonstrating that trust-based routing can deliver security without compromising responsiveness.
The flow installation time was measured at 4.5 ms, outperforming both centralised and distributed SDN architectures, such as those presented by Lee \textit{et al.}~\cite{Lee2023} (5.1 ms) and Li \textit{et al.}~\cite{Li2025} (5.8 ms). This efficiency arises from lightweight OpenFlow rule caching and a simplified trust-verification logic within the Ryu controller. Unlike Huang’s AI model~\cite{Huang2024}, which relies on iterative deep learning predictions before routing decisions, the proposed framework directly enforces trust-based flow rules, thereby reducing control-plane computation. The integration of a dual-channel architecture further minimises retransmission overhead during link switching, as evidenced by the reduced flow setup time at larger scales.

The average trust transition time of 1017~ms indicates rapid synchronisation between the Intrusion Detection System (IDS) and SDN controller. This is substantially lower than blockchain-driven or consensus-heavy systems such as those of Kumar \textit{et al.}~\cite{Kumar2023} (1290~ms) and Li \textit{et al.}~\cite{Li2025} (1120~ms). While federated learning-based IDS systems, such as those by Liu \textit{et al.}~\cite{Liu2025}, achieved comparable accuracy, they required additional computational time per node, raising concerns for deployment on resource-constrained UAVs. In contrast, the current design strikes a balance between trust recalibration and communication continuity, leveraging the fallback link to maintain operational stability even during reauthentication events.

In terms of reliability, the system maintained packet loss below 1\% over the primary (satellite/LTE) link and 3.5\% over the fallback (RF/mesh) link. This compares favourably with prior aerospace systems, including Huang’s AI-driven model~\cite{Huang2024} (1.8\%), Gao’s congestion-controlled networks~\cite{Gao2024} (1.6\%), and Singh’s multi-link UAV SDN setup~\cite{Singh2021} (2.2\%). The improvement is primarily due to real-time flow isolation of low-trust nodes, which prevents the propagation of malicious or degraded traffic through the network. Such adaptive control directly contributes to mission continuity and situational awareness, even in the presence of jamming or cyber interference.

Routing adaptability averaged 4.0~ms, outperforming adaptive learning-based or meta-heuristic frameworks such as those proposed by Alizadeh \textit{et al.}~\cite{Alizadeh2024} (5.2~ms) and Huang \textit{et al.}~\cite{Huang2024} (4.5~ms). As shown in Fig.~\ref{graph} and Table~\ref{tab:updated_graph_data}, scaling the network from 15 to 50 nodes resulted in only a slight increase in fallback delay (4.8 ms to 5.1 ms) and adaptability latency (4.0 ms to 4.5 ms), confirming excellent scalability. In larger topologies, flow installation times decreased marginally due to rule reuse, demonstrating improved efficiency with higher node density, an observation consistent with distributed SDN control research~\cite{Zhao2023, Lee2023}.

Overall, the proposed framework outperforms or matches recent aerospace communication architectures across all key parameters while introducing a unique combination of features not found collectively in any single prior work. These include real-time trust scoring integrated with zero-trust enforcement, centralised flow optimisation, and autonomous dual-channel failover. The sub-5~ms latency across multiple control operations demonstrates that trust-driven, software-defined networking can achieve both cybersecurity and real-time responsiveness, an essential requirement for UAV swarms, satellite coordination, and defence communication systems. Moreover, by maintaining performance consistency across network scales and link conditions, the architecture establishes a practical foundation for next-generation space–air–ground networks capable of resilient operation in contested or degraded environments.

\section{Conclusion}
This paper presents a trust-aware, dual-channel SDN framework for resilient UAV and aerospace communication networks, integrating a high-capacity primary link (satellite or LTE) with a low-power fallback channel (RF or mesh) and enforcing real-time zero-trust via an IDS-driven trust scoring mechanism. Experimental validation demonstrates sub-5~ms failover, efficient flow installation, and low packet loss, outperforming contemporary UAV SDN systems in both adaptability and trust responsiveness, confirming that SDN-based zero-trust architectures can substantially enhance cyber-physical resilience in next-generation aerospace and defence missions. Future work will enhance the trust-scoring mechanism through machine learning models that dynamically adjust thresholds beyond the current static, API-driven approach, and extend validation to hybrid simulation–hardware testbeds incorporating UAVs, satellites, and distributed SDN controllers. Additionally, integrating priority-based message transmission and emerging technologies such as 6G non-terrestrial networks and quantum-resistant cryptographic protocols will further harden the framework against sophisticated electronic and cyber warfare.

\section{Acknowledgements}
We would like to thank Mars Rover Manipal, an interdisciplinary student team of MAHE, for providing the resources needed for this project. WE also extend our gratitude to Dr Ujjwal Verma for his guidance and support in our work.

\bibliographystyle{IEEEtran}
\bibliography{main}

\end{document}